\documentclass[preprint, 11pt]{elsarticle}

\usepackage[T1]{fontenc}
\usepackage[utf8]{inputenc}
\usepackage[hmargin=0.9in]{geometry}
\usepackage[babel,german=quotes]{csquotes}
\usepackage{subcaption}
\usepackage{graphicx}
\usepackage[colorlinks=true,citecolor=blue,urlcolor=blue]{hyperref}
\usepackage{caption}
\usepackage{amsmath}
\usepackage{amsthm}
\usepackage{amssymb}
\usepackage{bm, amsfonts}
\usepackage{xcolor,soul,framed}
\usepackage{fixmath}
\usepackage{tikz}
\usetikzlibrary{patterns, arrows.meta}
\usepackage{pgfmath}
\usetikzlibrary{3d,shapes,positioning}

\usepackage{subcaption}
\usepackage{caption}

\usepackage{bm}
\usepackage{makecell}
\usepackage{titlesec}
\usepackage{multirow}
\usepackage{indentfirst}
\usepackage{mathtools}
\usepackage{float}
\usepackage[normalem]{ulem}
\useunder{\uline}{\ul}{}

\usepackage{booktabs} 
\usepackage{siunitx}  
\usepackage{adjustbox} 
\usepackage{tikz}

\colorlet{shadecolor}{green}

\titleformat{\paragraph}
{\normalfont\normalsize\itshape}{\theparagraph}{1em}{}
\titlespacing*{\paragraph}
{0pt}{3.25ex plus 1ex minus .2ex}{1.5ex plus .2ex}

\newdefinition{remark}{Remark}

\makeatletter
\def\ps@pprintTitle{%
\let\@oddhead\@empty
\let\@evenhead\@empty
\def\@oddfoot{
\footnotesize\itshape
\ifx\@journal\@empty Elsevier
\else\@journal\fi
\hfill\today
}%
\let\@evenfoot\@oddfoot}
\makeatother

\newcommand{\beq}{\begin{equation}}
\newcommand{\eeq}{\end{equation}}

\titleformat{\subsection}{\bfseries\small}{\thesubsection}{1em}{}

\begin{document}

\begin{frontmatter}
  \title{Effective Viscosity Closures for Dense Suspensions in CSP Systems via Lubrication-Enhanced DNS and Numerical Viscometry}
\author{Raphael Münster$^*$, Otto Mierka, Dmitri Kuzmin, Stefan Turek}
\ead{\{raphael.muenster,otto.mierka,dmitri.kuzmin,stefan.turek\}@math.tu-dortmund.de}
\cortext[cor1]{Corresponding author}

\address{Institute of Applied Mathematics (LS III), TU Dortmund University\\ Vogelpothsweg 87, D-44227 Dortmund, Germany}

\journal{}


\begin{abstract}
Dense particle suspensions are promising candidates for next-generation Concentrated Solar Power (CSP) receivers, enabling operating temperatures above 800\,\textdegree C. However, accurate modeling of the rheological behavior of granular flows is essential for reliable computational fluid dynamics (CFD) simulations. In this study, we develop and assess numerical methodologies for simulating dense suspensions pertinent to CSP applications.
Our computational framework is based on Direct Numerical Simulation (DNS), augmented by lubrication force models to resolve detailed particle-particle and particle-wall interactions at volume fractions exceeding 50\%.
We conducted a systematic series of simulations across a range of volume fractions to establish a robust reference dataset.
Validation was performed via a numerical viscometer configuration, permitting direct comparison with theoretical predictions and established benchmark results. Subsequently, the viscometer arrangement was generalized to a periodic cubic domain, serving as a representative volume element for CSP systems. Within this framework, effective viscosities were quantified independently through wall force measurements and energy dissipation analysis. The close agreement between these two approaches substantiates the reliability of the results.
Based on these findings, effective viscosity tables were constructed and fitted using polynomial and piecewise-smooth approximations. These high-accuracy closure relations are suitable for incorporation into large-scale, non-Newtonian CFD models for CSP plant design and analysis.
\end{abstract}

\begin{keyword}
Dense particle suspensions,
effective viscosity closure models,
lubrication-enhanced direct numerical simulation (DNS),
numerical viscometry,
non-Newtonian fluids,
concentrated solar power,
rheological modeling,
Euler-Euler methods,
high-performance CFD  
\end{keyword}

\end{frontmatter}

\section{Introduction}
\label{sec:intro}
Concentrated Solar Power (CSP) represents a critical pillar in the transition toward sustainable energy systems, offering not only the capability to convert solar radiation into electricity but also a practical solution for thermal energy storage. Unlike photovoltaic technologies that generate electricity directly from sunlight, CSP systems rely on optical concentrators to focus solar radiation onto a receiver, where the energy is absorbed and transformed into heat. This heat is then used to drive conventional power cycles, typically steam or gas turbines, for electricity production.

A key advantage of CSP lies in its intrinsic potential to store thermal energy efficiently, thereby decoupling energy collection from power generation and enabling dispatchable electricity supply during cloudy periods or after sunset.
Among various thermal storage media explored for CSP systems, dense particle suspensions have garnered significant attention as a next-generation heat transfer and storage medium. Comprising solid particles --- such as ceramic bauxite or sintered sand --- either suspended in a gas or used in pure particulate form, these suspensions exhibit superior thermal stability and energy density compared to traditional molten salts \cite{FLAMANT2013567}. More importantly, they enable operation at temperatures exceeding $800^\circ$C \cite{FLAMANT2014617}, a threshold that opens pathways to high-efficiency thermodynamic cycles such as the supercritical CO$_2$ Brayton cycle.

The deployment of dense suspensions in CSP introduces new modeling and engineering challenges. The inherently complex dynamics of dense granular flows---characterized by intense particle--particle and particle--fluid interactions---demand advanced numerical tools for analysis and system design. In this context, Computational Fluid Dynamics (CFD) has emerged as an indispensable tool, enabling the simulation of multiphase flows in CSP receivers, transport channels, and storage units. Through CFD, engineers can probe heat and momentum transfer mechanisms, assess flow stability, optimize geometry, and predict system performance under realistic conditions.
However, accurate numerical simulation of dense suspensions remains a formidable task. The constitutive behavior of such mixtures is strongly nonlinear and non-Newtonian, particularly at high particle concentrations. One of the central challenges is the proper modeling of the \emph{effective viscosity} of the suspension --- a macroscopic parameter that encapsulates the microscale rheological effects due to particle interactions and hydrodynamic forces. This viscosity depends sensitively on the local particle volume fraction, shear rate, and temperature, necessitating robust closure models for macroscopic-level CFD solvers.

Recent advances in CFD for disperse two-phase flows have furnished hybrid methodologies combining high-fidelity Direct Numerical Simulations (DNS) at the particle scale with coarse-grained Eulerian models. In particular, Discrete Network Approximations (DNA) and Arbitrary Lagrangian--Eulerian (ALE) finite element methods have shown promise in resolving the motion of rigid particles while retaining computational efficiency \cite{berlyand2005network,berlyand2005discrete,lefebvre2005apparent}. By performing extensive offline simulations across a range of operating conditions, researchers can derive polynomial or piecewise-smooth fits for the effective viscosity, thus supplying high-quality closures for use in large-scale non-Newtonian CFD models.

Technological advances in high-performance computing have significantly boosted the capabilities of traditional Euler--Euler multiphase flow methods.
The increased parallelism of modern hardware architectures, higher processor speeds, and the availability of massively parallel compute clusters have enabled Euler--Euler approaches to handle dense suspensions with unprecedented accuracy and efficiency. Today, these methods allow not only for stable and robust simulation of highly concentrated particle flows but also for the accurate computation of critical quantities such as effective viscosity, shear stress distributions, and particle volume fractions. This leap in computational feasibility positions Euler--Euler modeling strategies as a practical and reliable tool for the design, optimization, and operation of next-generation CSP plants. In this context, computational rheology presents both a compelling opportunity and a modeling frontier. The ability to understanding and predict the complex rheological behavior of  dense particle suspensions is essential for the design of efficient, reliable, and economically viable solar thermal power plants capable of meeting future energy demands.

To systematically address the above challenges, the remainder of this paper is structured as follows: Section~\ref{sec:modeling} formalizes the coupled fluid-particle model governing dense suspension flows, detailing the underlying physics and mathematical formulations. Sections~\ref{sec:fbm},~\ref{sec:fem} and~\ref{sec:lubrication} outline the numerical methodologies employed, including the finite element discretization and the implementation of lubrication force models. Validation of the computational framework is presented in Section~\ref{sec:viscometry}, where simulation results are compared against theoretical predictions and benchmark data. In Subsection~\ref{subsec:cubic_rve}, we generalize the viscometer configuration to a periodic domain to derive effective viscosity closures pertinent to CSP applications. Finally, Section~\ref{sec:conclusion} summarizes the key findings and discusses their implications for large-scale CFD simulations in CSP plant design.

\section{Physical and Mathematical Modeling}
\label{sec:modeling}
Adopting the Finite Element Method–Fictitious Boundary Method (FEM–FBM) approach to direct numerical simulation \cite{WanTurek2006a}, we consider a coupled system of governing equations for the fluid and $N$ solid particles. These equations are solved in the fictitious domain 
\[
\Omega_T = \Omega_f \cup \bigcup_{i=1}^N \Omega_i,
\]
where $\Omega_f$ denotes the subdomain occupied by the fluid and $\Omega_i,\ i\in\{1,\ldots,N\}$ is the subdomain occupied by the $i$-th particle.

\subsection{Fluid Flow}
The motion of the incompressible fluid is governed by the \textbf{Navier–Stokes equations}
\begin{align}
\rho_f \left( \frac{\partial \mathbf{u}}{\partial t} + \mathbf{u} \cdot \nabla \mathbf{u} \right) - \nabla \cdot \boldsymbol{\sigma} &= \mathbf{0}, & \text{(Momentum equation)} \\
\nabla \cdot \mathbf{u} &= 0, & \text{(Continuity equation)}
\end{align}
where $\mathbf{u}$ is the fluid velocity field, $\rho_f$ is the fluid density, and $\boldsymbol{\sigma}$ is the Cauchy stress tensor. These equations are valid in the fluid domain $\Omega_f(t)$, which depends on the time $t$. The Cauchy stress tensor for a Newtonian fluid is defined as
\begin{equation}
\boldsymbol{\sigma} = -p I + \mu_f \left[\nabla \mathbf{u} + (\nabla \mathbf{u})^T\right].
\end{equation}
Here, $p$ is the pressure, $I$ is the identity tensor, and $\mu_f$ is the fluid dynamic viscosity. The corresponding kinematic viscosity is given by $\nu=\mu_f/\rho_f$.

\subsection{Particle Motion}
Each rigid particle $\Omega_i$ undergoes translational and rotational motion governed by the Newton–Euler equations. Let $\mathbf{U}_i(t)$ and $\boldsymbol{\omega}_i(t)$ denote the translational and angular velocities of the $i$-th particle, respectively. The equations of motion read
\begin{subequations}\label{newtoneuler}
\begin{align}
M_i \frac{d \mathbf{U}_i}{dt} &= \Delta M_i \mathbf{g} + \mathbf{F}_i + \mathbf{F}_i^{\text{col}}, \\
\mathbf{I}_i \frac{d \boldsymbol{\omega}_i}{dt} + \boldsymbol{\omega}_i \times (\mathbf{I}_i \boldsymbol{\omega}_i) &= \mathbf{T}_i,
\end{align}
\end{subequations}
where
\begin{itemize}
  \item $M_i$ is the mass of the $i$-th particle,
  \item $\mathbf{I}_i$ is the moment of inertia tensor,
  \item $\Delta M_i = M_i - \rho_f \lvert \Omega_i \rvert$ is the effective buoyant mass,
  \item $\mathbf{g}$ is the gravitational acceleration,
  \item $\mathbf{F}_i$ and $\mathbf{T}_i$ are the hydrodynamic force and torque acting on the particle,
  \item $\mathbf{F}_i^{\text{col}}$ accounts for collisions with particles or walls and includes forces that model close-range lubrication effects. 
\end{itemize}

The hydrodynamic force and torque are computed from the stress distribution on the particle boundary $\partial \Omega_i$ as follows:
\begin{align}
\mathbf{F}_i &= -\int_{\partial \Omega_i} \boldsymbol{\sigma} \cdot \mathbf{n} \, d\Gamma, \\
\mathbf{T}_i &= -\int_{\partial \Omega_i} (\mathbf{x} - \mathbf{X}_i) \times (\boldsymbol{\sigma} \cdot \mathbf{n}) \, d\Gamma,
\end{align}
where $\mathbf{X}_i$ is the center of mass and $\mathbf{n}$ denotes the unit outward normal to $\partial \Omega_i$.

The particle’s position and orientation are determined by the ordinary differential equations
\begin{align}
\frac{d \mathbf{X}_i}{dt} &= \mathbf{U}_i, \\
\frac{d \boldsymbol{\theta}_i}{dt} &= \boldsymbol{\omega}_i,
\end{align}
where $\boldsymbol{\theta}_i$ denotes the rotational state, e.g., the rotation vector depending on the chosen representation.

On the fluid–particle interface $\partial \Omega_i$, the velocities satisfy the no-slip boundary condition 
\begin{equation}\label{rbm}
\mathbf{u}(\mathbf{x}) = \mathbf{U}_i + \boldsymbol{\omega}_i \times (\mathbf{x} - \mathbf{X}_i) \quad \forall \mathbf{x} \in \partial\Omega_i
\end{equation}
corresponding to the rigid body motion in $\Omega_i$.


\section{Multigrid FEM Fictitious Boundary Method}
\label{sec:fbm}

The \emph{Multigrid Finite Element Method–Fictitious Boundary Method} (FEM–FBM) provides an efficient approach for simulating particulate flows by avoiding remeshing and enabling parallel scalability \cite{WanTurek2006a, WanTurek2006b}. The central idea is to run simulations on a single, fixed background mesh that covers the entire computational domain $\Omega_T = \Omega_f \cup \bigcup_{i=1}^N \Omega_i$. The method is closely related to the fictitious domain approaches presented in  \cite{glowinski1,glowinski2,patankar2000}.

\subsection{Fictitious Domain Formulation}

The FEM-FBM methodology extends the Navier–Stokes equations to the full domain $\Omega_T$, incorporating the rigid particle constraint via an explicit enforcement of the no-slip condition within the particle regions. The governing equations are formulated as:
\begin{subequations}
\label{eq:fbm_system}
\begin{align}
\nabla \cdot \mathbf{u} &= 0, && \text{in } \Omega_T, \\
\rho_f \left( \frac{\partial \mathbf{u}}{\partial t} + \mathbf{u} \cdot \nabla \mathbf{u} \right) - \nabla \cdot \boldsymbol{\sigma} &= \mathbf{0}, && \text{in } \Omega_f, \\
\mathbf{u}(\mathbf{x}) &= \mathbf{U}_i + \boldsymbol{\omega}_i \times (\mathbf{x} - \mathbf{X}_i), && \text{in } \overline{\Omega}_i,\quad i = 1,\dots,N.
\end{align}
\end{subequations}

Here, the constraint \eqref{eq:fbm_system}c enforces rigid body motion inside each particle domain $\Omega_i$, effectively transforming the problem into a single-domain formulation with implicitly embedded moving obstacles.

\subsection{Hydrodynamic Force Calculation via Volume Integration}

To avoid reconstructing particle surfaces at each time step, the FBM computes hydrodynamic forces and torques via a volume-integral approximation. Define the indicator function $\alpha_i : \Omega_T \to \{0,1\}$ by
\[
\alpha_i(\mathbf{x}) = 
\begin{cases}
1, & \text{if } \mathbf{x} \in \Omega_i, \\
0, & \text{otherwise}.
\end{cases}
\]
The interface is implicitly represented through the gradient $\nabla \alpha_i$, which is non-zero only near $\partial \Omega_i$. The hydrodynamic force $\mathbf{F}_i$ and torque $\mathbf{T}_i$ are then computed as
\begin{align}
\mathbf{F}_i &= -\int_{\Omega_T} \boldsymbol{\sigma} \cdot \nabla \alpha_i \, d\mathbf{x}, \\
\mathbf{T}_i &= -\int_{\Omega_T} (\mathbf{x} - \mathbf{X}_i) \times (\boldsymbol{\sigma} \cdot \nabla \alpha_i) \, d\mathbf{x}.
\end{align}
This volume-based formulation is particularly well suited for structured meshes, as it requires integration only over a narrow band of cells surrounding each particle.

\subsection{Numerical Strategy}

The multigrid FEM–FBM algorithm advances the fluid–solid mixture in time using the following steps at each time level:

\begin{enumerate}
  \item Solve the fluid equations \eqref{eq:fbm_system}a and \eqref{eq:fbm_system}b over $\Omega_T$, imposing the rigid-body constraints via \eqref{eq:fbm_system}c.
  \item Compute the hydrodynamic forces and torques using the representation as volume integrals.
  \item Update translational and angular velocities of the particles by solving the Newton–Euler equations \eqref{newtoneuler}.
  \item Update particle positions and apply new rigid-body constraints to the fluid field.
\end{enumerate}

The background grid remains fixed, and particles move freely through it. The multigrid solver efficiently handles the pressure–velocity coupling and nonlinearities introduced by convection, maintaining optimal performance even for large numbers of particles.

\section{Finite Element Discretization}
\label{sec:fem}
\subsection{Time Integration: Fractional-Step-$\theta$ Scheme}

The coupled fluid–solid system is discretized in time using the \textit{fractional-step-$\theta$ scheme} \cite{Blasco_Codina_Huerta_1998, WanTurek2007a}, which is strongly A-stable and particularly effective for stiff, transient flow problems. The scheme introduces minimal numerical dissipation and supports accurate resolution of oscillatory behavior.

Let $\mathbf{u}^n$ and $p^n$ denote the fluid velocity and pressure at time $t^n$, and let $K = t^{n+1} - t^n$ denote the time step size. In
our description of the time-stepping method, we will use the operator notation
\[
\mathcal{N}(\mathbf{v}) \mathbf{u} := -\nu \nabla \cdot \left( \nabla \mathbf{u} + \nabla \mathbf{u}^\top \right) + \mathbf{v} \cdot \nabla \mathbf{u}
\]
for the sum of the convective and viscous terms in the discretized momentum equations.

The update from $t^n$ to $t^{n+1}$ is performed in three substeps. The degree of
implicitness for individual steps is determined by the parameters
\[
\theta = 1 - \frac{\sqrt{2}}{2}, \qquad \theta' = 1 - 2\theta, \qquad \alpha = \frac{1 - 2\theta}{1 - \theta}, \qquad \beta = 1 - \alpha.
\]
The first two steps of the following algorithm
produce intermediate velocities $\mathbf{u}^{n+\theta}$ and $\mathbf{u}^{n+1-\theta}$. The third
step yields the final velocity $\mathbf{u}^{n+1}$.

\subsection*{Step 1: First Intermediate Velocity}
\begin{align}
\left[ I + \alpha \theta K\, \mathcal{N}(\mathbf{u}^{n+\theta}) \right] \mathbf{u}^{n+\theta} + \theta K \nabla p^{n+\theta} &= \left[ I - \beta \theta K\, \mathcal{N}(\mathbf{u}^n) \right] \mathbf{u}^n, \\
\nabla \cdot \mathbf{u}^{n+\theta} &= 0.
\end{align}

\subsection*{Step 2: Second Intermediate Velocity}
\begin{align}
\left[ I + \beta \theta' K\, \mathcal{N}(\mathbf{u}^{n+1-\theta}) \right] \mathbf{u}^{n+1-\theta} + \theta' K \nabla p^{n+1-\theta} &= \left[ I - \alpha \theta' K\, \mathcal{N}(\mathbf{u}^{n+\theta}) \right] \mathbf{u}^{n+\theta}, \\
\nabla \cdot \mathbf{u}^{n+1-\theta} &= 0.
\end{align}

\subsection*{Step 3: Final Velocity Update}
\begin{align}
\left[ I + \alpha \theta K\, \mathcal{N}(\mathbf{u}^{n+1}) \right] \mathbf{u}^{n+1} + \theta K \nabla p^{n+1} &= \left[ I - \beta \theta K\, \mathcal{N}(\mathbf{u}^{n+1-\theta}) \right] \mathbf{u}^{n+1-\theta}, \\
\nabla \cdot \mathbf{u}^{n+1} &= 0.
\end{align}

\subsection*{Rigid Body Motion Constraint}
After each velocity update, the velocity of the fictitious fluid is constrained to satisfy the constraint 
\begin{equation}\label{rbm2}
\mathbf{u}^{n+1}(\mathbf{x}) = \mathbf{U}_i^{n+1} + \boldsymbol{\omega}_i^{n+1} \times \left( \mathbf{x} - \mathbf{X}_i^{n+1} \right), \qquad \forall\, \mathbf{x} \in \overline{\Omega}_i
\end{equation}
of rigid body motion in the particle domain $\Omega_i$. In our FEM-FBM method, this
constraint is imposed in a strong sense by extending
the Dirichlet boundary condition \eqref{rbm} into the interior $\Omega_i$ of the particle.
At the fully discrete level, condition
\eqref{rbm2} is enforced at the nodal points
that belong to $\bar\Omega_i$.
This approach eliminates the need for using
body-fitted grids or distributed Lagrange 
multipliers.

\subsection*{A Note on Implementation}
Each substep results in a nonlinear saddle-point problem that can be solved using fixed-point iteration or linearization (e.g., Oseen approximation). The pressure updates may be performed using discrete projection methods and multigrid solvers~\cite{Turek1999}.


\section*{Space Discretization with the \( Q_2/P^{\mathrm{disc}}_{1} \) Element Pair on Hexahedral Meshes}

We discretize the incompressible Navier–Stokes equations in three spatial dimensions using a background mesh of conforming hexahedra.

\subsection*{Finite Element Spaces}

Let \( \mathcal{T}_h \) denote a shape‑regular family of hexahedral elements covering the extended computational domain \( \Omega_T \subset \mathbb{R}^3 \). We define the discrete velocity and pressure spaces as:
\[
\begin{aligned}
H_h^{Q_2} &:= \left\{ \mathbf{v}_h \in C^0(\Omega_T)^3 \; \middle| \;
                 \mathbf{v}_h|_{T} \in Q_2(T)^3 \;\; \forall T \in \mathcal{T}_h,\;
                 \mathbf{v}_h = \mathbf{0} \text{ on } \partial\Omega_T \right\}, \\[2pt]
L_h^{P^{\mathrm{disc}}_1} &:= \left\{ q_h \in L^2_0(\Omega_T) \; \middle| \;
                 q_h|_{T} \in P_1(T) \;\; \forall T \in \mathcal{T}_h\;
                 \right\},
\end{aligned}
\]
where \( Q_2(T) \) denotes the space of triquadratic polynomials and \( P_1(T) \) the space of linear polynomials on each element \( T \) of the mesh.

\subsection*{Bilinear and Trilinear Forms}

For \( \mathbf{u}_h, \mathbf{v}_h, \mathbf{w}_h \in H_h^{Q_2} \) and \( p_h \in L_h^{P^{\mathrm{disc}}_1} \), we define
\[
\begin{aligned}
a_h(\mathbf{u}_h, \mathbf{v}_h)        &:= \sum_{T \in \mathcal{T}_h}
                        \int_T \nabla \mathbf{u}_h : \nabla \mathbf{v}_h \,dx, \\
b_h(p_h, \mathbf{v}_h)                &:= -\sum_{T \in \mathcal{T}_h}
                        \int_T p_h \, (\nabla \cdot \mathbf{v}_h) \, dx, \\
n_h(\mathbf{u}_h, \mathbf{v}_h, \mathbf{w}_h) &:= \sum_{T \in \mathcal{T}_h}
                        \int_T \left[ (\mathbf{u}_h \cdot \nabla) \mathbf{v}_h \right] \cdot \mathbf{w}_h \, dx.
\end{aligned}
\]

\subsection*{Streamline-Diffusion Stabilization}

To stabilize the convective terms, we use the modified trilinear form
\[
\tilde{n}_h(\mathbf{u}_h, \mathbf{v}_h, \mathbf{w}_h)
    := n_h(\mathbf{u}_h, \mathbf{v}_h, \mathbf{w}_h)
     + \sum_{T \in \mathcal{T}_h}
       \delta_T \int_T (\mathbf{u}_h \cdot \nabla \mathbf{v}_h)(\mathbf{u}_h \cdot \nabla \mathbf{w}_h) \, dx,
\]
with the local stabilization parameter
\[
\delta_T := \delta^\ast
            \frac{h_T}{\|\mathbf{u}\|_{\Omega_T}}
            \cdot \frac{2\, \mathrm{Re}_T}{1 + \mathrm{Re}_T}, \qquad
\mathrm{Re}_T := \frac{\|\mathbf{u}\|_T \cdot h_T}{\nu},
\]
where \( h_T \) is the element diameter, \( \|\mathbf{u}\|_T \) is a local norm (e.g., \( L^2 \)), and \( \delta^\ast \in [0.1, 1] \) is a user-chosen constant.

\subsection*{Fully Discrete Saddle Point Problems}

In each stage of the fully discrete fractional-step method for the Navier-Stokes system, we compute a finite element approximation \( (\mathbf{u}_h, p_h) \in H_h^{Q_2} \times L_h^{P^{\mathrm{disc}}_1} \) such that
\[
\begin{aligned}
(\mathbf{u}_h, \mathbf{v}_h)
&+ \theta_1 K \left[ a_h(\mathbf{u}_h, \mathbf{v}_h) + \tilde{n}_h(\mathbf{u}_h, \mathbf{u}_h, \mathbf{v}_h) \right]
+ \theta_2 K \, b_h(p_h, \mathbf{v}_h)
   = (\mathbf{f}, \mathbf{v}_h)
& \quad \forall\, \mathbf{v}_h \in H_h^{Q_2}, \\[4pt]
b_h(q_h, \mathbf{u}_h) &= 0
& \quad \forall\, q_h \in L_h^{P^{\mathrm{disc}}_1}.
\end{aligned}
\]
The use of discrete projection methods as iterative solvers for the corresponding nonlinear saddle-point problems makes it possible to update the velocity and pressure in a segregated manner \cite{Turek1999}.

\subsection*{Fictitious Boundary Conditions}

Given the coordinates of the center of mass
\( \mathbf{X}_i(t) \), the translational velocity \( \mathbf{U}_i(t) \), and angular velocity \( \boldsymbol{\omega}_i(t) \) of
a rigid particle \( \Omega_i(t) \), we
constrain the velocity degrees of freedom at all nodal points $x_{\mathrm{DOF}}\in\bar\Omega_i(t)$ to satisfy the fictitious boundary conditions
\[
\mathbf{u}_h(x_{\mathrm{DOF}})
   := \mathbf{U}_i(t) + \boldsymbol{\omega}_i(t) \times \left( x_{\mathrm{DOF}} - \mathbf{X}_i(t) \right),
\qquad x_{\mathrm{DOF}} \in \bar\Omega_i(t).
\]
That is, we treat $x_{\mathrm{DOF}}$ in the same way as a Dirichlet boundary node.

\subsection*{Hydrodynamic Force Evaluation}

The total hydrodynamic force acting on particle \( i \) is computed via
\[
\mathbf{F}_i := -\int_{\Omega_T}
        \boldsymbol{\sigma}(\mathbf{u}_h, p_h) : \nabla \alpha_i \, dx,
\qquad
\alpha_i(x) =
\begin{cases}
1, & x \in \Omega_i(t), \\
0, & \text{otherwise},
\end{cases}
\]
where the Cauchy stress tensor is defined as
\[
\boldsymbol{\sigma}(\mathbf{u}_h, p_h)
     = -p_h \mathbf{I} + 2\nu \mathbf{D}(\mathbf{u}_h), \qquad
\mathbf{D}(\mathbf{u}_h) := \frac{1}{2} \left( \nabla \mathbf{u}_h + \nabla \mathbf{u}_h^\top \right).
\]
For the details of the solution scheme and efficiency considerations, we refer to \cite{Turek1999,afc3}.

\section{Hard-Sphere Particle Model and Lubrication Forces}
\label{sec:lubrication}

\subsection{Contact Formulation}
\label{subsec:contact_model}

We use a semi-implicit time-stepping method for frictional contact between rigid bodies, combining contact detection, impulse computation, and constraint projection based on the formulations by Stewart \& Trinkle \cite{stewart_trinkle_1996} and Anitescu \& Potra \cite{anitescu_potra_1997}. The employed framework supports both sequential and distributed execution.
Contacts between rigid bodies are modeled as point-wise constraints using a local frame \(\{\bm{n}, \bm{t}, \bm{o}\}\) aligned with the normal \(\bm{n}\) and two tangential directions. The computation of contact impulses \(\bm{p}_i = (p_n, p_t, p_o)\) ensures that non-penetration and Coulomb friction are enforced:
\[
\dot{\bm{g}}_i = \bm{W}_i \bm{p}_i + \bm{b}_i, \quad \bm{p}_i \in \mathcal{K}_i,
\]
where \(\bm{W}_i\) is the Delassus operator (or Schur complement), and \(\mathcal{K}_i\) is the local friction cone: \(\|\bm{p}_t\| \leq \mu p_n,\, p_n \geq 0\).

The normal component of the velocity is stabilized using the Baumgarte correction \cite{baumgarte_1972}, which modifies the target velocity to counteract constraint drift.

\subsection{Solver Structure}

We employ two contact solvers:
\begin{itemize}
    \item \textbf{Approximate Decoupled Solver}: Updates normal and tangential impulses separately (as in \cite{hwangbo_lee_hutter_2018}). Normal impulses are projected onto the positive half-line; tangential ones are projected into a disk of radius \(\mu p_n\).
    \item \textbf{Orthogonal Projection Solver}: Based on Anitescu \& Tasora \cite{anitescu_tasora_2010}, updates the full impulse and projects the result onto the friction cone. This retains coupling between normal and tangential impulses and improves robustness.
\end{itemize}

Both solvers operate within an iterative loop, updating one contact at a time and accumulating the resulting velocity changes. This structure naturally leads to the \textbf{Projected Gauss–Seidel (PGS)} method \cite{erleben_2004, wang2015warmstartingprojectedgaussseidel}.

\subsection{Projected Gauss–Seidel (PGS)}

The PGS algorithm is an iterative solver for contact impulses:
\[
\bm{p}_i^{k+1} = \Pi_{\mathcal{K}_i} \left( \bm{p}_i^k - \omega \bm{W}_{ii}^{-1} \dot{\bm{g}}_i \right),
\]
where \(\Pi_{\mathcal{K}_i}\) is the projection onto the admissible set (e.g., friction cone) and \(\omega \in (0,1]\) is a relaxation parameter.

Unlike matrix solvers, this method loops over contacts, immediately updating the bodies’ velocities. This ``local and sequential'' structure is what makes it a Gauss–Seidel variant — but adapted to inequality-constrained dynamics.

\subsection*{Parallelization with Domain Decomposition}

Although PGS is inherently a sequential algorithm in the sense that each contact update depends on the latest values of neighboring contacts, it can be effectively used in parallel simulations via domain decomposition.
In a distributed contact solver, the simulation domain is partitioned into subdomains assigned to different processors. Bodies that lie near domain boundaries are duplicated as \emph{ghost bodies} (or shadow copies) on neighboring subdomains. Each domain then proceeds with its own local PGS loop over its contact set.
Consistency across domains is maintained as follows:
\begin{itemize}
    \item After each full PGS iteration (i.e., one sweep over all contacts), each subdomain synchronizes the velocities of ghost bodies with their host domain.
    \item This synchronization ensures that when a neighboring domain continues its sweep, it sees the latest velocity updates from contacts affecting shared bodies.
    \item Communication can be implemented via MPI or other inter-process protocols and typically involves only boundary data.
\end{itemize}

This simple synchronization strategy allows the PGS solver to operate in a Gauss–Seidel fashion \emph{within each domain}, while enabling globally consistent evolution of contact dynamics across all domains. In this way the CFD solver and the Particle solver can both run in parallel using the same domain decomposition, which makes this approach an excellent fit for our purposes.

\subsection{Lubrication Forces}
\label{subsec:lubrication_forces}

The lubrication model that we use for near-contact interactions in viscous suspensions is compatible with rigid body solvers based on hard contact formulations. It follows the approach of Kroupa et al.~\cite{Kroupa_Kosek_Soos_Vonka_2016} and Dance \& Maxey~\cite{DANCE2003212,YEO20102401}, adapted to match the variable conventions of the hard contact solver. 

The variables and symbols that appear below in the formulas for individual forces are specific to bilateral particle interactions. For a complete explanation of these quantities, we refer to \ref{appendix:notation}, which summarizes their definitions and the context.

\subsection*{Normal Force}
\begin{equation}
F_{n}^{\text{lub}} = -6\pi \mu_f R_p \dot{g}_n \left( \frac{1}{\epsilon} - \frac{9}{40} \log \epsilon - \frac{3}{112} \epsilon \log \epsilon \right).
\end{equation}

\subsection*{Tangential (Sliding) Force}
\begin{equation}
\bm{F}_{t}^{\text{lub}} = -6\pi \mu_f R_p \left[ \bm{v}_t \left(-\frac{1}{6} \log \epsilon\right) + \bm{v}_{ct} \left(-\frac{1}{6} \log \epsilon - \frac{1}{12} \epsilon \log \epsilon\right) \right].
\end{equation}

\subsection*{Sliding Torque}
\begin{align}
\bm{M}_{t}^{\text{lub}} = -8\pi \mu_f R_p^2 \bigg[ &(\bm{n} \times \bm{v}_t) \left(-\frac{1}{6} \log \epsilon - \frac{1}{12} \epsilon \log \epsilon \right) \nonumber \\
&+ (\bm{n} \times \bm{v}_{ct}) \left(-\frac{1}{5} \log \epsilon - \frac{47}{250} \epsilon \log \epsilon \right) \bigg].
\end{align}

\subsection*{Twisting Torque}
\begin{equation}
\bm{M}_{n}^{\text{lub}} = -8\pi \mu_f R_p^2 \left[ (\bm{\omega}_i - \bm{\omega}_j) \cdot \bm{n} \right] \bm{n} \left( \frac{1}{8} \epsilon \log \epsilon \right).
\end{equation}

\subsection*{Slip Regularization}

To avoid singularity as \(\epsilon \to 0\), we
apply the slip correction factor 
\begin{equation}
f^* = \frac{h}{3h_c} \left[ \left(1 + \frac{h}{6h_c} \right) \ln \left(1 + \frac{6h_c}{h} \right) - 1 \right]
\end{equation}
from \cite{Vinogradova_Yakubov_2003}.
The corrected lubrication force and torque are given by
\[
\bm{F}^{\text{lub, corrected}} = f^* \bm{F}^{\text{lub}}, \quad \bm{M}^{\text{lub, corrected}} = f^* \bm{M}^{\text{lub}}.
\]
If \(h < h_c\), we set \(h = h_c\) and \(\epsilon = \epsilon_c = h_c / R_p\) beforehand.

\subsection*{Particle-Wall Interactions}

Interactions with solid walls are modeled using the same formulas with \(\bm{v}_j = \bm{0}, \bm{\omega}_j = \bm{0}\), and \(h\) being 
the distance between
the particle surface and the wall. We use a slight tweak of the fractional terms, as outlined in \cite{Kroupa_Kosek_Soos_Vonka_2016}.

\section{Numerical Viscometry and Validation Framework}
\label{sec:viscometry}

\subsection{Numerical Rotational Viscometer}
\label{subsec:rotational_viscometer}

In order to characterize the rheological properties of particle-laden incompressible flows, we simulate a classical concentric cylinder (Couette-type) rotational viscometer. This configuration is widely used as a benchmark for measuring the effective viscosity of suspensions at various volume fractions of the particulate phase. A brief description of the experimental/computational setup is given below. The details can be found in the publication of Prignitz and Baensch~\cite{faucris.109633524}.

\subsubsection*{Geometry and Operating Principle}

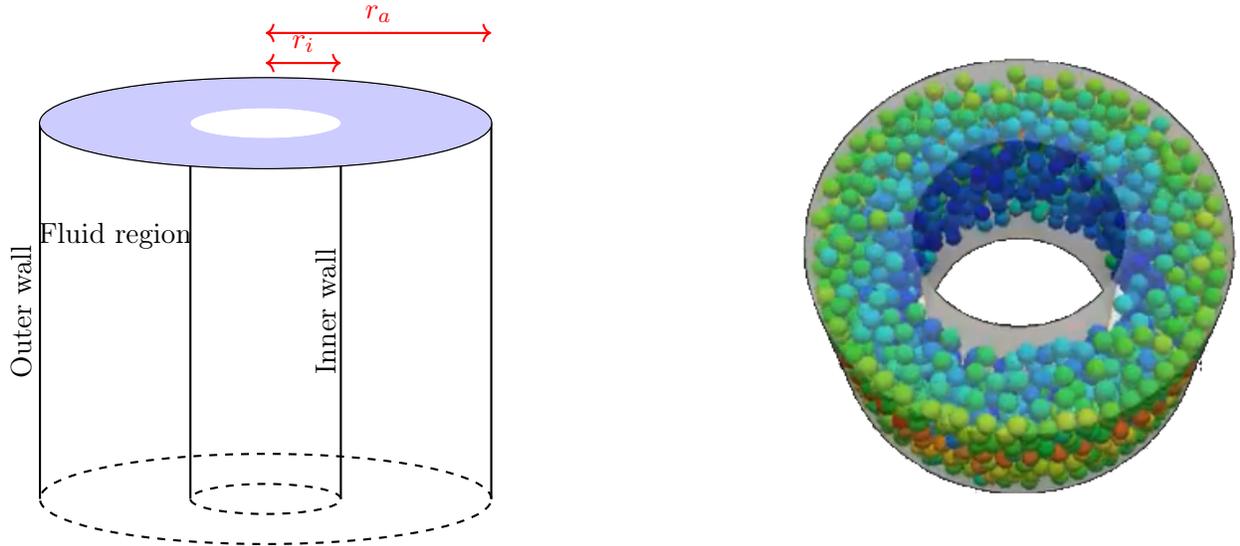
\begin{figure}[H]
  \centering
  \begin{subfigure}[t]{0.4\textwidth}
    \centering  
\begin{tikzpicture}

  \draw[thick] (0,0) ellipse (3 and 0.6);

  \draw[thick] (-3,0) -- (-3,-5);
  \draw[thick] (3,0) -- (3,-5);

  \draw[thick, dashed] (0,-5) ellipse (3 and 0.6);

  \draw[thick] (0,0) ellipse (1 and 0.2);

  \draw[thick] (-1,0) -- (-1,-5);
  \draw[thick] (1,0) -- (1,-5);

  \draw[thick, dashed] (0,-5) ellipse (1 and 0.2);

  \begin{scope}
    \clip (0,0) ellipse (3 and 0.6);
    \fill[blue!20] (0,0) ellipse (3 and 0.6);
  \end{scope}
  \begin{scope}
    \clip (0,0) ellipse (1 and 0.2);
    \fill[white] (0,0) ellipse (1 and 0.2);
  \end{scope}

  \node at (-2, -1.5) {Fluid region};
  \node[rotate=90] at (-3.25,-2.5) {Outer wall};
  \node[rotate=90] at (0.8,-2.5) {Inner wall};

  \draw[<->, thick, red] (0,0.8) -- (1,0.8) node[midway, above] {$r_i$}; 
  \draw[<->, thick, red] (0,1.2) -- (3,1.2) node[midway, above] {$r_a$}; 

\end{tikzpicture}

  \end{subfigure}
  \hfill
  \begin{subfigure}[t]{0.4\textwidth}
    \centering
    \includegraphics[width=\linewidth]{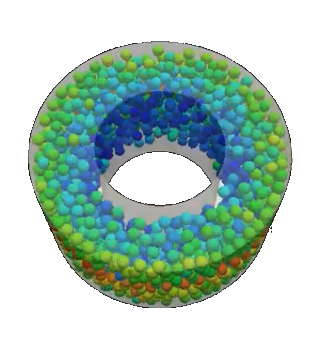}
  \end{subfigure}

  \caption{Experimental setup and simulation with particle volume fraction of 35\% in the Numerical Viscometer.}
  \label{fig:comparison}
\end{figure}

The simulated device consists of two concentric cylinders: the outer cylinder (referred to as the \emph{cup}) is set into rotation with angular velocity $\omega$, while the inner cylinder (referred to as the \emph{bob}) remains stationary. The annular gap between the two cylinders is filled with either a Newtonian reference fluid or a dense suspension. The torque exerted by the fluid on the inner cylinder provides a measure of the suspension’s effective viscosity. A sketch of the experimental setup is shown in Figure~\ref{fig:comparison}.

\subsubsection*{Shear Rate and Shear Stress}

Under the assumption of steady, axisymmetric flow in the narrow-gap limit, the azimuthal velocity profile is approximately linear. The shear rate $\dot{\gamma}$ at the outer wall is approximated by
\begin{equation}
    \dot{\gamma} = \frac{du_\phi}{dr} \approx \frac{2\pi r_a \omega}{r_a - r_i}, 
    \label{eq:shear_rate}
\end{equation}
where $r_i$ and $r_a$ denote the inner and outer cylinder radii, respectively.

The corresponding wall shear stress $\tau$ can be related to the measured torque $T$ via
\begin{equation}
    \tau = \frac{T}{2\pi r_a^2}.
    \label{eq:shear_stress}
\end{equation}

\subsubsection*{Effective Viscosity Estimation}

Using the definitions above, the dynamic viscosity $\mu$ of the fluid or suspension is computed from the measured torque as
\begin{equation}
    \mu = \frac{T (r_a - r_i)}{4\pi^2 \omega r_a^3}.
    \label{eq:viscosity}
\end{equation}
This expression confirms that the viscosity is directly proportional to the measured torque for fixed geometry and rotation rate.

\subsubsection*{Relative and Effective Viscosity}

To assess the impact of suspended particles on the rheology, we define the dimensionless \emph{relative viscosity} $\mu^*$ as the ratio of the effective viscosity of the suspension $\mu_{\text{eff}}$ to that of the pure fluid $\mu$:
\begin{equation}
    \mu^* = \frac{\mu_{\text{eff}}}{\mu} = \frac{T_{\text{eff}}}{T},
    \label{eq:relative_viscosity}
\end{equation}
where $T_{\text{eff}}$ is the torque measured for the suspension and $T$ is the reference torque obtained for the Newtonian fluid. In the numerical implementation, $T_{\text{eff}}$ is computed by integrating the torque contribution from hydrodynamic stresses over the surface of the outer cylinder. Using formula \eqref{eq:relative_viscosity}, we calculated the effective viscosity $\mu_{\text{eff}}$ corresponding to the results of our numerical torque measurements. 
\medskip

\begin{figure}[htbp]
  \centering
  
  \begin{subfigure}[t]{\textwidth}
   \centering
    \includegraphics[width=0.7\linewidth]{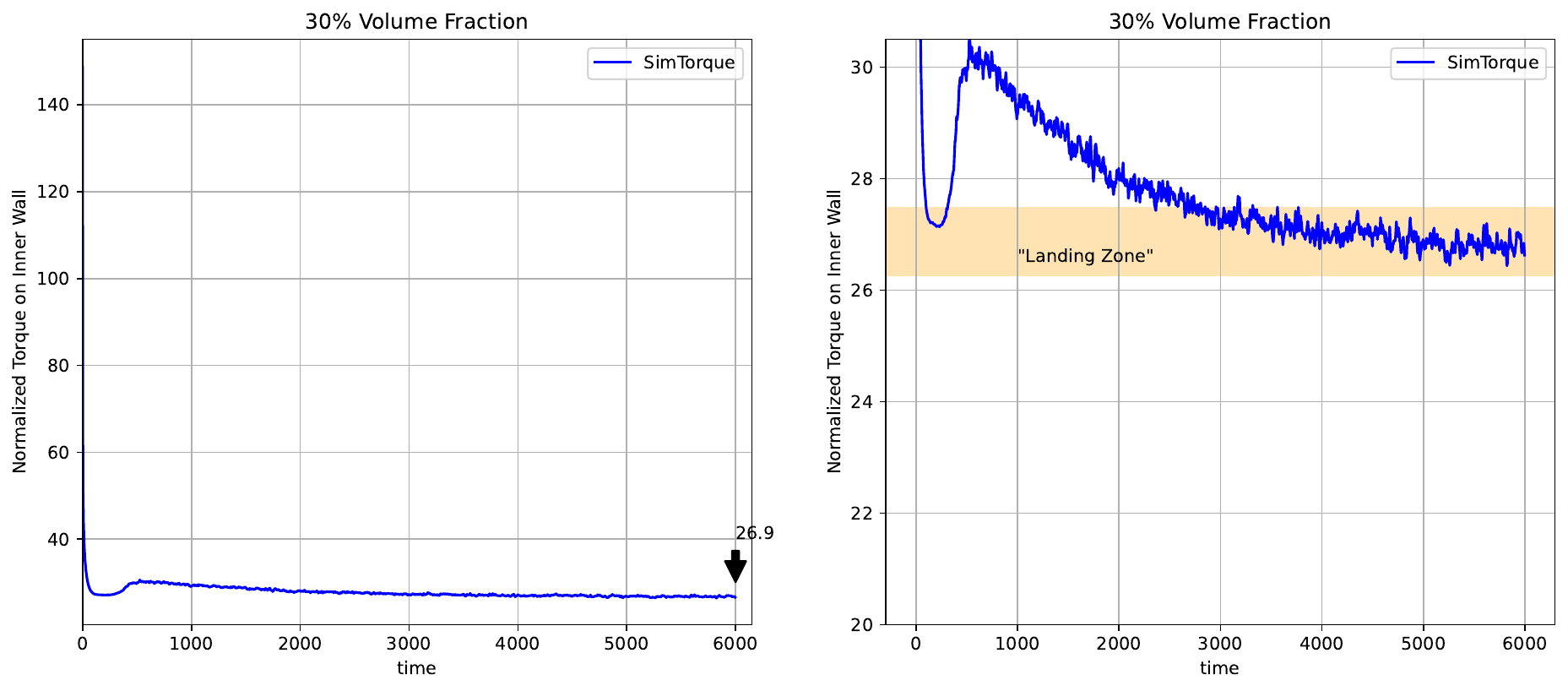}
    \caption{Convergence history for 30\%.}
    \label{fig:plot1}
  \end{subfigure}
  
  \vspace{1em} 
  \begin{subfigure}[t]{\textwidth}
    \centering
    \includegraphics[width=0.7\linewidth]{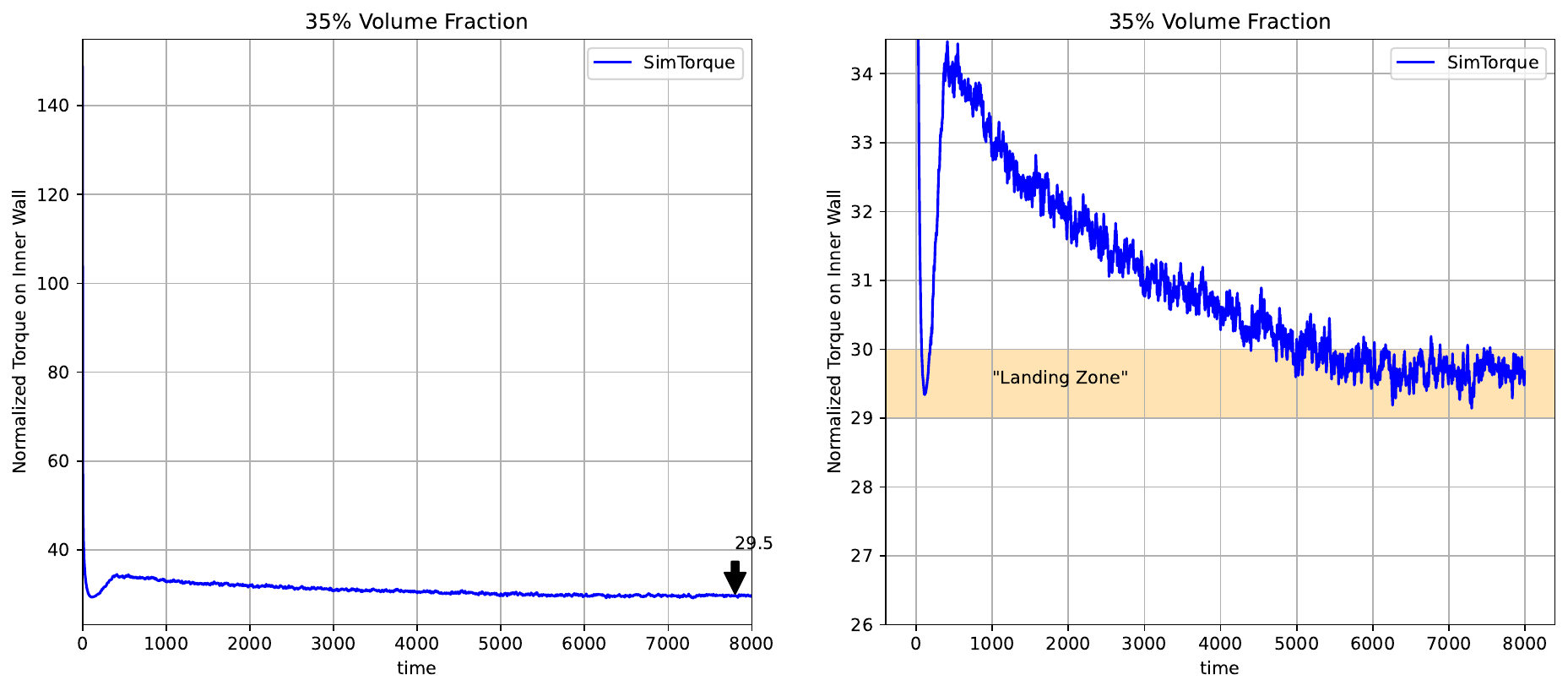}
    \caption{Convergence history for 35\%.}
    \label{fig:plot4}
  \end{subfigure}
  
  \vspace{1em} 
  \begin{subfigure}[t]{\textwidth}
    \centering
    \includegraphics[width=0.7\linewidth]{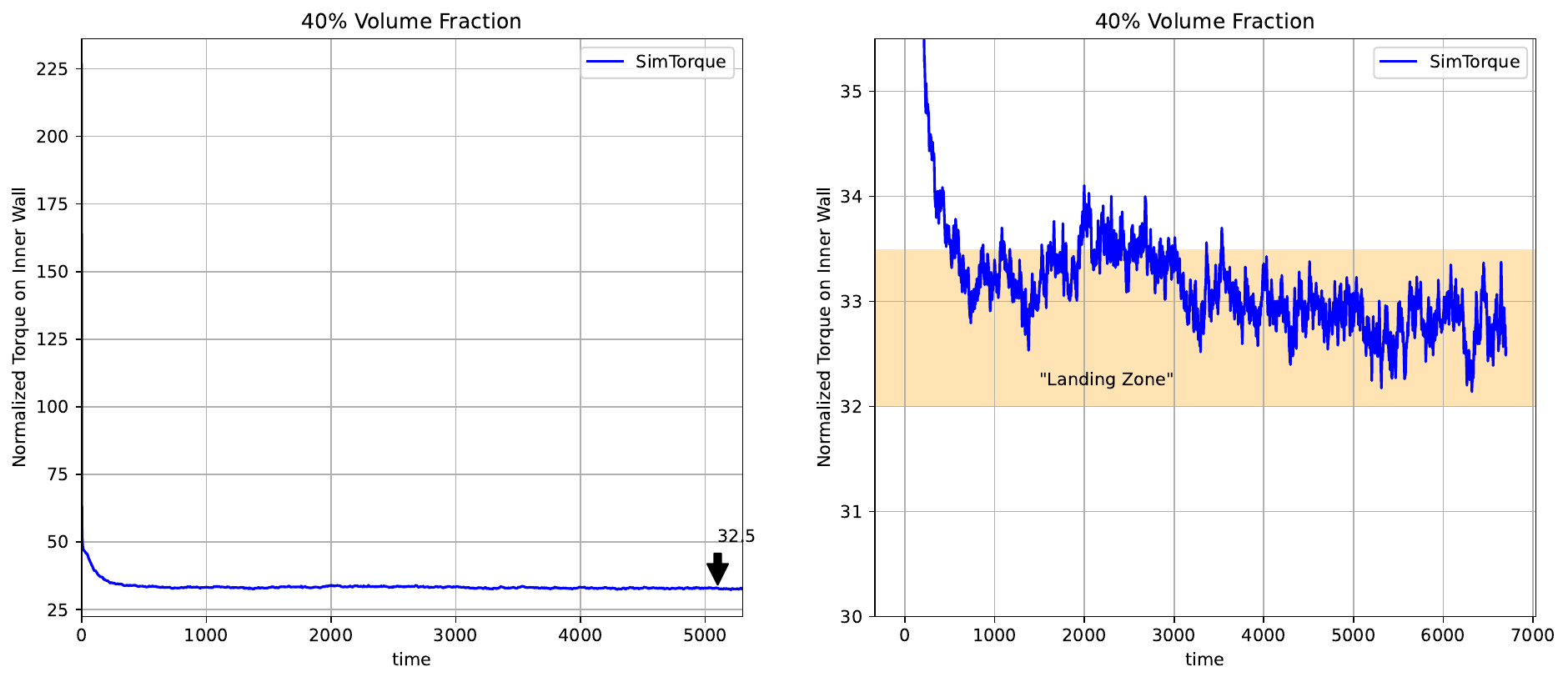}
    \caption{Convergence history for 40\%.}
    \label{fig:plot4}
  \end{subfigure}

  \caption{Time evolution of the torque measurements for volume fractions of
  30\%, 35\% and 40\%.}
  \label{fig:all_plots}
\end{figure}

We observe an interesting behavior of our FEM-FBM approach to numerical viscometry in Fig. \ref{fig:all_plots}: In the initial stage of the simulation, a large overshoot (compared to the final value) is followed by a sharp drop. Then the measured torque increases again before beginning to settle and converge with minor oscillations around the mean in a ``landing zone''. We take this mean as the final value. This behavior of the method is also visible in our computations in a periodic cubic domain. 

\begin{figure}[H]
    \centering
    \includegraphics[width=0.8\textwidth]{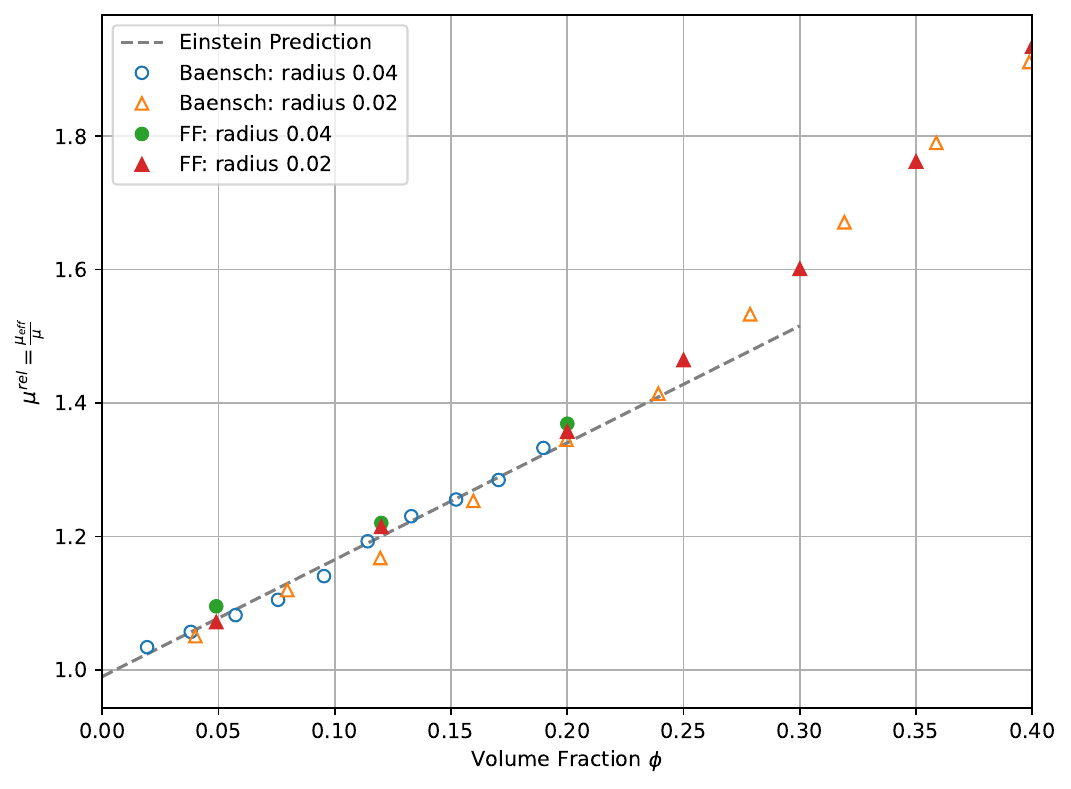} 
    \caption{Numerical Viscometer, 3D simulation results obtained with the  FEM-FBM method (FF) vs. Einstein's
    effective viscosity formula and 2D 
    simulation results from \cite{faucris.109633524}.}
    \label{fig:num_visc_results}
\end{figure}

In Fig. \ref{fig:num_visc_results}, we plot the results of our numerical viscometer simulation vs. effective viscosities computed by  Prignitz and Baensch~\cite{faucris.109633524} using a similar fictitious domain method. Additionally, we show the effective viscosities calculated using Einstein's formula. We see that for relatively low volume fractions, the effective viscosity increases linearly in accordance with Einstein's theory. Another observation is that the results of both numerical simulations start to deviate from the linear behavior for volume fractions around 20-25\%. This confirms that dense suspensions of particles behave as non-Newtonian fluids. Remarkably, our simulations produce the same results for particle of different radii, but with the same volume fraction. In summary, the presented FEM-FBM results are in very good agreement with existing theory and reference data. In particular, the 
response of the effective viscosity to an increase in the volume fraction of particles is captured correctly by our 3D simulation.

\subsection{Cubic RVE for Dense Suspensions}
\label{subsec:cubic_rve}
In this section we validate the core component of our effective viscosity toolbox: measuring the effective viscosity in a cubic RVE using DNS. 
\subsection*{Validation Using Wall Force Balance Method}

In addition to the energy-dissipation based evaluation of $\mu_{\text{eff}}$ on the periodic RVE domain, we cross-validated our results by implementing a wall-driven shear configuration as presented in Kroupa et al.~\cite{Kroupa_Kosek_Soos_Vonka_2016}. This approach computes the effective viscosity via direct force balance on a moving wall in a simple shear cell.

\subsection*{Computational Setup} The domain is a cube of size $L \times L \times L$ with periodic boundary conditions in the $x$ and $z$ directions. The bottom wall is fixed, while the top wall moves at a constant velocity $U_x$, thereby imposing a known shear rate $\dot{\gamma} = U_x / L$. The suspension is sheared between these walls, and the top wall simultaneously acts as a probe surface for measuring the drag force. 

\subsection*{Force Decomposition} The total shear stress $\tau$ acting on the upper wall is determined as
\begin{equation}
\tau = \frac{F_{\text{drag},x}}{A},
\end{equation}
where $F_{\text{drag},x}$ is the $x$-component of the total drag force exerted by the suspension on the upper wall and $A = L^2$ is the wall area.

This drag force comprises two contributions:
\begin{itemize}
  \item The hydrodynamic force $F_{H,x}$, obtained by integrating the fluid stress $\mu_f \partial u_x / \partial y$ at $y=L$;
  \item The particle-wall lubrication force $F_{L,x}$, evaluated by summing lubrication interactions from particles in near-contact with the top wall.
\end{itemize}

The corresponding components of the effective viscosity are given by
\begin{equation}\label{munew}
\mu_H = \frac{F_{H,x}}{A \dot{\gamma}}, \quad
\mu_L = \frac{F_{L,x}}{A \dot{\gamma}}, \quad
\mu_{\text{eff}} = \mu_H + \mu_L.
\end{equation}

\subsection*{Interpretation and Consistency}
\begin{figure}[h]
    \centering
    \includegraphics[width=0.6\textwidth]{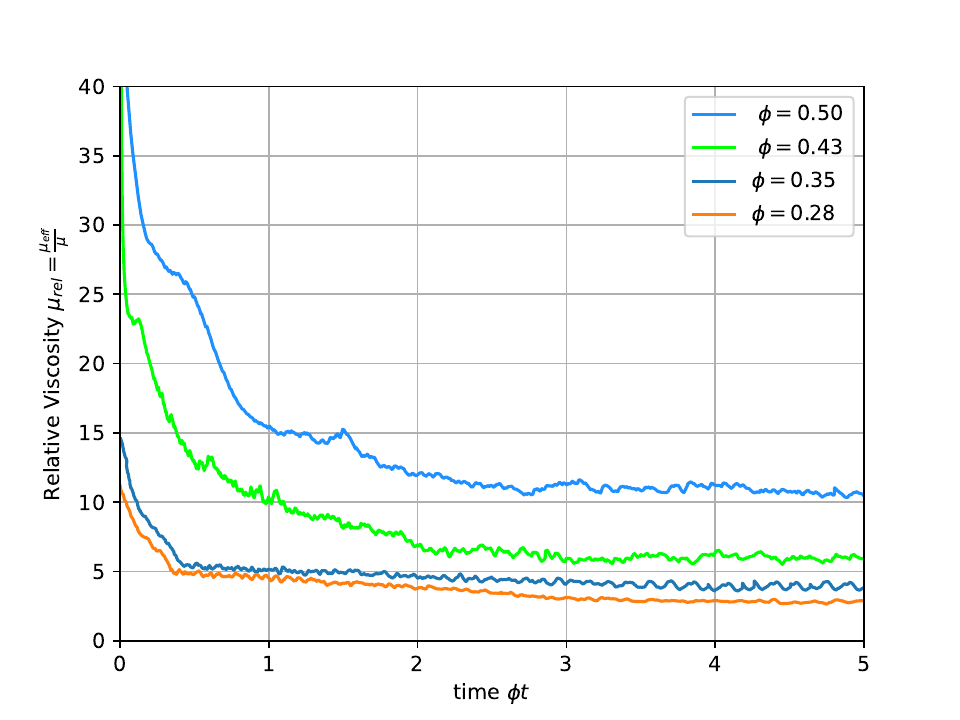} 
    \caption{Time evolution of the relative viscosity.}
    \label{fig:kroupa_results}
\end{figure}

Formula \eqref {munew} yields a fully mechanical definition of
$\mu_{\text{eff}}$  consistent with Newton’s law of viscosity
\[
\tau = \mu_{\text{eff}} \dot{\gamma}
\]
and provides an alternative to the energy-based assessment that we used to estimate the effective viscosity in our RVE analysis.
We verified that for the same particle configuration, the two approaches yield effective viscosities in excellent agreement. Moreover, both formulations exhibit the expected overshoot–relaxation behavior during start-up shear before stabilizing in a quasi-steady ``landing zone'', as seen in Fig.~\ref{fig:kroupa_results} and the results in~\cite{Kroupa_Kosek_Soos_Vonka_2016}.

For further validation, we evaluated the effective viscosity using a definition that compares the dissipation in the particle-laden suspension to that of a reference shear flow:

\[
\mu_{\text{eff}} = \mu\,
\frac{
\displaystyle \int_{\Omega_{\text{sus}}} D(u) : D(u)\,dx
}{
\displaystyle \int_{\Omega_{\text{coarse}}} D(\bar u) : D(\bar u)\,dx
},
\]
where $D(u) = \tfrac{1}{2} \left( \nabla u + \nabla u^\top \right)$ denotes the strain-rate tensor, and the numerator measures the total viscous dissipation in the suspension domain $\Omega_{\text{sus}}$. The denominator represents the dissipation in a coarse-scale reference flow field $\bar u$ that would arise in a pure fluid domain with the same shear rate $\dot\gamma$.
For the special case of a cubic RVE under simple shear, this reference dissipation reduces to

\[
\int_{\Omega_{\text{coarse}}} D(\bar u) : D(\bar u)\,dx = \tfrac{1}{2} \dot\gamma^2\,|\Omega|,
\]
which corresponds to the dissipation in a uniform Newtonian fluid of viscosity $\mu$ sheared at constant rate $\dot\gamma$ across the volume $\Omega$. This expression serves as a normalization factor, allowing the ratio to quantify how much the presence of particles increases the energy dissipation — and thereby the apparent viscosity — relative to the pure fluid.

\begin{figure}[h]
    \centering
    \includegraphics[width=0.75\textwidth]{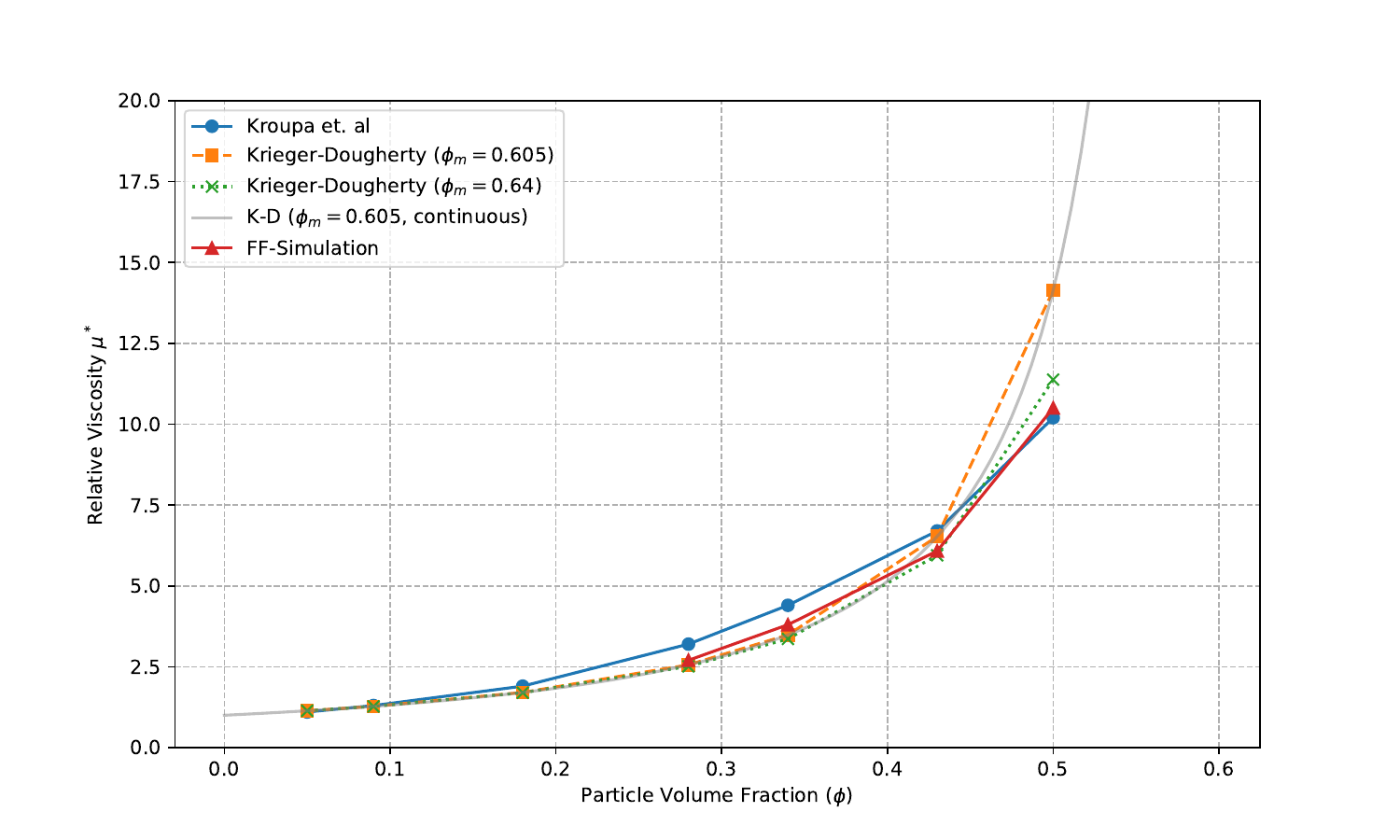} 
    \caption{Plots corresponding to the values in Table \ref{tab:rel_visc_krieger_comparison}.}
    \label{fig:kroupa_kd_comp}
\end{figure}

\begin{figure}[h]
    \centering
    \includegraphics[width=0.7\textwidth]{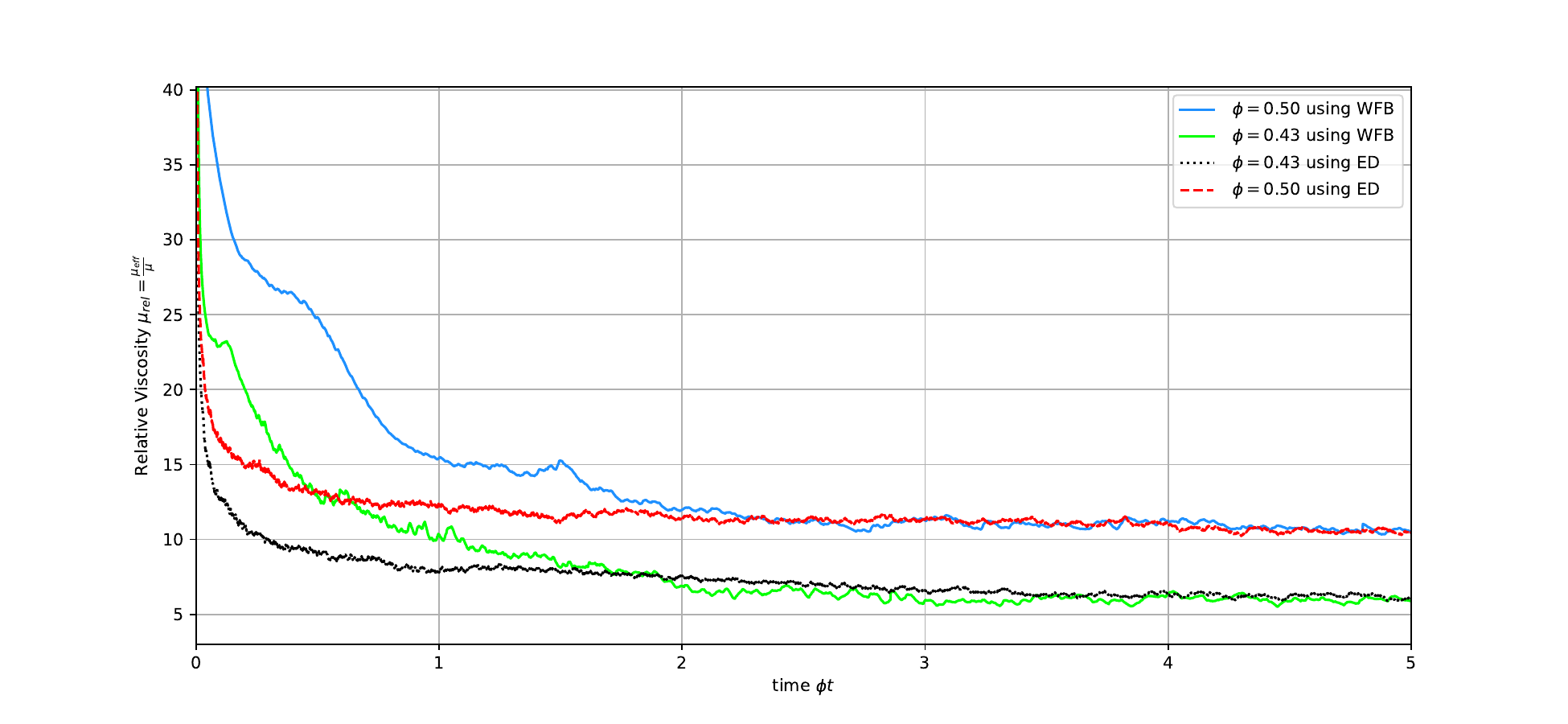} 
    \caption{Comparison of effective viscosity calculation using energy dissipation (ED) and wall force balance (WFB).}
    \label{fig:kroupa_ed_comp}
\end{figure}

Table \ref{tab:rel_visc_krieger_comparison} and Fig. \ref{fig:kroupa_kd_comp} show how well our results and those obtained by Kroupa et al. \cite{Kroupa_Kosek_Soos_Vonka_2016} agree with the Krieger–Dougherty (K--D) model predictions at different solid volume fractions $\phi$. In Fig.~\ref{fig:kroupa_ed_comp}, we compare the effective viscosities produced by the energy dissipation (ED) approach and the results obtained using wall force balancing (WFB) method.  Excellent agreement is observed between the two predictions. Besides giving further evidence of the correct setup of our RVE measurements, this also tells us that our DNS resolution is sufficient to capture the steep velocity gradients between closely spaced particles, which is the expected dominant source of viscous dissipation in this configuration.
This test confirms the robustness of our lubrication-augmented hard-contact model and the consistency of both techniques for calibration and validation of coarse-grained viscosity closures.

\begin{table}[h]
\centering
\caption{Comparison of relative viscosities $\mu^* = \mu_s / \mu_f$ from simulations and Krieger–Dougherty (K--D) model predictions at different solid volume fractions $\phi$. Two variants of the K--D model are considered:  $\phi_m^{\mathrm{kd}}$ and the random close packing $\phi_m^{\mathrm{kd,rcp}}$. We compare our results (FF) with the simulations of Kroupa et al. \cite{Kroupa_Kosek_Soos_Vonka_2016}.}
\begin{tabular}{@{}c
                S[table-format=2.2]
                S[table-format=2.2]
                S[table-format=2.2]
                S[table-format=2.2]
                @{}}
\toprule
{$\phi$}
& {Kroupa et al.}
& {FF}
& {K--D $\mu^*$ $(\phi_m = \phi_m^{kd})$} 
& {K--D $\mu^*$ $(\phi_m = \phi_m^{kd,rcp})$} \\
\midrule
0.28 & 3.20 & 2.70 & 2.56 & 2.51 \\
0.34 & 4.40 & 3.80 & 3.49 & 3.36 \\
0.43 & 6.70 & 6.08 & 6.53 & 5.95 \\
0.50 & 10.20 & 10.50 & 14.14 & 11.38 \\
\bottomrule
\end{tabular}
\label{tab:rel_visc_krieger_comparison}
\end{table}

\subsection{Effective Viscosity Closures}
\label{subsec:closures}

The final component of our numerical toolbox for effective viscosity evaluations
is a polynomial fit obtained from steady‑state RVE simulations that cover the desired input space sufficiently. We then fit a bivariate polynomial of the form
\begin{equation}
\mu_{\mathrm{eff}}(\alpha, \dot\gamma) =
\sum_{k=0}^{n_k} \sum_{l=0}^{n_l} c_{kl} \, \alpha^k \, \dot\gamma^l,
\label{eq:closure}
\end{equation}
where $c_{kl}$ are the coefficients to be determined by least-squares fitting.
This expression represents a tensor-product polynomial basis
of degree $n_k$ in $\alpha$ and $n_l$ in $\dot\gamma$.
For numerical efficiency, the polynomial can be evaluated using
the nested Horner scheme
\begin{align}
P_k(\dot\gamma) &= \sum_{l=0}^{n_l} c_{kl} \, \dot\gamma^l,
\\
\mu_{\mathrm{eff}}(\alpha, \dot\gamma) &= \sum_{k=0}^{n_k} \alpha^k \, P_k(\dot\gamma).
\end{align}
The fitting using the above scheme can be done using $Re_p$ instead of $\dot{\gamma}$ to combine different carriers inside the plateau regime as well as data points that are outside of this regime.

In order to align the RVE to the operating conditions of CSP systems, we choose $\dot{\gamma}$ in the range of $10-400 \; [s^{-1}]$. Furthermore, we take an air carrier fluid with $\mu = 0.0185 \; [\text{mPa s}]$, which corresponds to the CSP setup in \cite{articleAnsartRenaud}. We include a water-based carrier and an intermediate carrier for comparison purposes. In the targeted configuration the particle Reynolds number 
\[
Re_p = \frac{\rho_f R^{2}_p \dot{\gamma}}{\mu}
\]
remained well below 0.1.   
In this creeping flow regime, the effective viscosity becomes independent of shear rate, resulting in a viscosity plateau that depends solely on the solid volume fraction $\alpha$. This behavior is consistent with findings in the literature ~\cite{PhysRevLett.129.078001, PhysRevLett.109.118305}.

To validate our closure model, we conducted simulations across various carrier fluids (air, intermediate viscosity fluids, and water) and solid volume fractions (30\%, 40\%, 45\%, and 50\%). The results, summarized in Table~\ref{tab:viscosity-summary}, confirm that under CSP operating conditions, the flow remains within the shear-rate-independent viscosity plateau. This validation supports the applicability of our polynomial closure model for predicting effective viscosity in macroscopic simulations of CSP systems. 

\begin{table}[ht]
\centering
\caption{Summary of effective viscosity results across carrier fluids, volume fractions $\phi$, and regimes (plateau vs. inertial).}
\begin{tabular}{@{}ll
                S[table-format=2.0, round-mode=none]    
                S[table-format=1.2, round-mode=none]    
                S[table-format=1.3, round-mode=none]    
                S[table-format=2.3, round-mode=none]    
                S[table-format=2.2, round-mode=none]    
                S[table-format=3.0, round-mode=none]    
                @{\hspace{1.25em}}}
\toprule
\textbf{Carrier} & \textbf{Case} & {$\phi$ (\%)} & {$d$ (mm)} & {$\mathrm{Re}_p$} & {$\mu_{\mathrm{eff}}$ (mPa·s)} & {$\mu_r$} & {$\dot{\gamma}$} \\
\midrule
\multirow{3}{*}{\parbox{2cm}{Air \\ $\mu=0.018$, $\rho=1.2$}} 
& A0 & 30 & 0.1 & 0.0074 & 0.054 & 3.01 & 50 \\
& A1 & 40 & 0.1 & 0.0074 & 0.096 & 5.38 & 50 \\
& A2 & 45 & 0.1 & 0.0074 & 0.134 & 7.47 & 50 \\
& A3 & 50 & 0.1 & 0.0074 & 0.197 & 10.95 & 50 \\
\midrule
\multirow{3}{*}{\parbox{2cm}{Intermed. \\ $\mu=0.10$, $\rho=1000$}} 
& I0 & 30 & 0.224 & 0.0125 & 0.554 & 5.54 & 100 \\
& I1 & 40 & 0.224 & 0.0125 & 0.554 & 5.54 & 100 \\
& I2 & 45 & 0.224 & 0.0125 & 0.788 & 7.88 & 100 \\
& I3 & 50 & 0.224 & 0.0125 & 1.114 & 11.14 & 100 \\
\midrule
\multirow{5}{*}{\parbox{2cm}{Water \\ $\mu=1.00$, $\rho=1000$}} 
& W0 & 30 & 0.04 & 0.040 & 2.86 & 2.86 & 100 \\
& W1 & 40 & 0.04 & 0.040 & 5.24 & 5.24 & 100 \\
& W2 & 45 & 0.04 & 0.040 & 7.32 & 7.32 & 100 \\
& W3 & 50 & 0.04 & 0.040 & 10.51 & 10.51 & 100 \\
& W4 & 40 & 0.04 & 0.12 & 5.54 & 5.54 & 300 \\
& W5 & 40 & 0.04 & 0.16 & 5.65 & 5.65 & 400 \\
& W6 & 40 & 0.04 & 0.20 & 5.79 & 5.79 & 500 \\
\bottomrule
\end{tabular}
\label{tab:viscosity-summary}
\end{table}

\begin{figure}[h]
    \centering
    \includegraphics[width=0.7\textwidth]{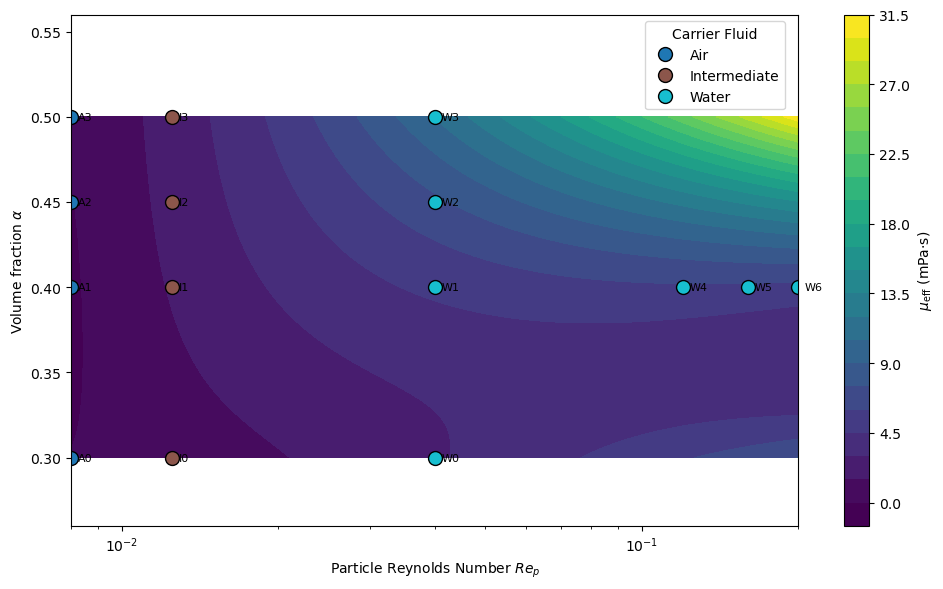} 
    \caption{Fitted closure $\mu_{\mathrm{eff}}(\alpha, \log_{10}(Re_p))$
    based on steady-state RVE simulations.}
    \label{fig:num_visc_results2}
\end{figure}
Figure~\ref{fig:num_visc_results2} shows the effective viscosity maps generated from our simulations in Table \ref{tab:viscosity-summary}, providing a comprehensive view of the viscosity behavior across different operating conditions. These maps serve as a valuable resource for predicting flow behavior in CSP systems and related applications. With the help of the proposed scheme, custom maps can be generated for different operating conditions.

\subsection*{Mild Inertia Regime and Viscosity Correction Model}
The data points W4-W6 of Table \ref{tab:viscosity-summary} show that we have actually gradually left the viscosity plateau regime in which the effective viscosity $\mu_{\text{eff}}$ of a dense suspension is independent of the imposed shear rate $\dot\gamma$, and depends solely on the particle volume fraction $\phi$. As $Re_p$ increases toward $\mathcal{O}(0.1)$, small but finite inertial effects emerge. This regime is called the \emph{mild inertia regime}, characterized by deviations from the Stokes solution while the bulk flow remains laminar. These effects are due to fluid inertia at the particle scale.
To quantify the rise in viscosity in this regime, Boyer et al.~\cite{PhysRevLett.107.188301} and Trulsson et al.~\cite{PhysRevLett.109.118305} proposed a first-order correction to the relative viscosity:
\[
\mu_r^{\text{eff}}(\phi, \mathrm{Re}_p) \;=\; \mu_r(\phi) \left(1 + \frac{C(\phi)}{\mu_r(\phi)}\,\mathrm{Re}_p \right),
\]
or, equivalently,
\[
\mu_{\text{eff}} = \mu_f\,\mu_r(\phi)\left(1 + \frac{C(\phi)}{\mu_r(\phi)}\,\mathrm{Re}_p\right).
\]
The empirical coefficient $C(\phi)$ depends on the volume fraction $\phi$ and encapsulates the strength of inertial microstructural interactions.
The increase in relative viscosity is then expressed as:
\[
\Delta := \frac{\mu_r^{\text{eff}}}{\mu_r} - 1 = \frac{C(\phi)}{\mu_r(\phi)}\,\mathrm{Re}_p.
\]
This relationship is \emph{linear in $\mathrm{Re}_p$} for small-to-moderate values, typically $\mathrm{Re}_p \lesssim 0.6$.

In the work of Boyer et al.~\cite{PhysRevLett.107.188301} and Trulsson et al.~\cite{PhysRevLett.109.118305}, we find that for $\phi\in[0.4,0.5]$, DNS and experiments yield $C(\phi) \in [1.5, 3.0]$, depending on the definition of $\mu_r$, surface friction, and the $\mathrm{Re}_p$ fitting window. The mild inertia model is validated up to $\mathrm{Re}_p \approx 0.6$, beyond which nonlinear effects begin to dominate, and the stress scales more closely as $\dot\gamma^2$.

From our data points W4-W6, we can conclude that our DNS approach can capture the mild inertia regime, and the data suggests that for our approach $C(0.4)$ is 2.5$\sim$2.7. This is consistent with observations of Boyer and Trulsson that there can be a statistical noise and micro-structural differences between runs that contribute to a ±5–10\% variance.

\section{Conclusion}
\label{sec:conclusion}
We have developed and validated an accurate DNS-based toolbox for numerical viscometry experiments using a finite element fictitious boundary method. The objective of the presented numerical study was to derive fitted effective viscosity closures for coarse-grained simulations of dense particle suspensions in CSP systems. The results have been validated using different carriers, domain geometries, and evaluation techniques. The employed DNS simulation tool is able to capture phenomena like viscosity plateaus and predict the effects of onsetting inertia in the mild inertial regime. 

  \section*{Acknowledgments}
  This work was supported by the German Research Foundation (DFG) under grant
  KU 1530/28-1 (TU 102/77-1). The authors gratefully acknowledge collaboration
   on this project
  with Prof. Yuliya Gorb (National Science Foundation) and Prof. Alexey Novikov
  (Pennsylvania State University).

\appendix
\section{Notation and Variable Definitions}
\label{appendix:notation}

This appendix summarizes the key symbols and expressions used in the lubrication force model (Section~\ref{subsec:lubrication_forces}) for bilateral interactions in viscous suspensions. These definitions align with the hard-contact solver framework and are valid for both particle-particle and particle-wall interactions.

\begin{itemize}
  \item \(\mu_f\): Dynamic viscosity of the fluid.
  \item \(R_p\): Radius of a spherical particle.
  \item \(h\): Surface-to-surface distance between particle centers minus \(2R_p\), \(h = |\bm{x}_i - \bm{x}_j| - 2R_p\).
  \item \(\epsilon = h / R_p\): Dimensionless surface gap.
  \item \(\bm{n}\): Unit normal vector pointing from particle \(j\) to particle \(i\).
  \item \(\bm{v}_r = \bm{v}_i - \bm{v}_j\): Relative translational velocity between particles.
  \item \(\dot{g}_n = \bm{v}_r \cdot \bm{n}\): Normal component of relative velocity (positive for separating motion).
  \item \(\bm{v}_t = \bm{v}_r - \dot{g}_n \bm{n}\): Tangential sliding velocity.
  \item \(\bm{\omega}_i, \bm{\omega}_j\): Angular velocity vectors of particles \(i\) and \(j\).
  \item \(\bm{r}_k = R_p \bm{n}\): Contact point vector on particle \(k \in \{i,j\}\).
  \item \(\bm{v}_{c,k} = \bm{\omega}_k \times \bm{r}_k\): Surface velocity at the contact point due to rotation.
  \item \(\bm{v}_{ct} = \bm{v}_{c,i} - \bm{v}_{c,j}\): Tangential slip due to differential rotation.
  \item \(\log \epsilon\): Natural logarithm of the dimensionless gap, appearing in asymptotic expansions of hydrodynamic forces.
\end{itemize}

\bibliographystyle{plain} 
\bibliography{references} 

\begin{thebibliography}{10}

\bibitem{anitescu_potra_1997}
Mihai Anitescu and Florian~A. Potra.
\newblock Formulating dynamic multi-rigid-body contact problems with friction
  as solvable linear complementarity problems.
\newblock {\em Nonlinear Dynamics}, 14(3):231--247, 1997.

\bibitem{anitescu_tasora_2010}
Mihai Anitescu and Alessandro Tasora.
\newblock An iterative approach for cone complementarity problems for nonsmooth
  dynamics.
\newblock {\em Computational Optimization and Applications}, 47(2):207--235,
  2010.

\bibitem{articleAnsartRenaud}
Renaud Ansart, Pablo García~Triñanes, Benjamin Boissiere, Hadrien Benoit,
  J.P.K. Seville, and Olivier Simonin.
\newblock Dense gas-particle suspension upward flow used as heat transfer fluid
  in solar receiver: Pept experiments and 3d numerical simulations.
\newblock {\em Powder Technology}, 307, 11 2016.

\bibitem{baumgarte_1972}
J.~Baumgarte.
\newblock Stabilization of constraints and integrals of motion in dynamical
  systems.
\newblock {\em Computer Methods in Applied Mechanics and Engineering},
  1(1):1--16, 1972.

\bibitem{berlyand2005network}
Leonid Berlyand, Liliana Borcea, and Alexander Panchenko.
\newblock Network approximation for effective viscosity of concentrated
  suspensions with complex geometry.
\newblock {\em SIAM Journal on Mathematical Analysis}, 36(5):1580--1628, 2005.

\bibitem{berlyand2005discrete}
Leonid Berlyand, Yuliya Gorb, and Alexei Novikov.
\newblock Discrete network approximation for highly-packed composites with
  irregular geometry in three dimensions.
\newblock In {\em Multiscale Methods in Science and Engineering}, pages 21--57.
  Springer, 2005.

\bibitem{Blasco_Codina_Huerta_1998}
J.~Blasco, R.~Codina, and A.~Huerta.
\newblock A fractional-step method for the incompressible {N}avier-{S}tokes
  equations related to a predictor-multicorrector algorithm.
\newblock {\em International Journal for Numerical Methods in Fluids},
  28(10):1391–1419, Dec 1998.

\bibitem{PhysRevLett.107.188301}
Fran\ifmmode \mbox{\c{c}}\else~\c{c}\fi{}ois Boyer, \'Elisabeth Guazzelli, and
  Olivier Pouliquen.
\newblock Unifying suspension and granular rheology.
\newblock {\em Phys. Rev. Lett.}, 107:188301, Oct 2011.

\bibitem{DANCE2003212}
S.L. Dance and M.R. Maxey.
\newblock Incorporation of lubrication effects into the force-coupling method
  for particulate two-phase flow.
\newblock {\em Journal of Computational Physics}, 189(1):212--238, 2003.

\bibitem{erleben_2004}
Kenny Erleben.
\newblock {\em Stable, Robust, and Versatile Multibody Dynamics Animation}.
\newblock PhD thesis, University of Copenhagen, 2004.

\bibitem{FLAMANT2014617}
G.~Flamant, D.~Gauthier, H.~Benoit, J.-L. Sans, B.~Boissière, R.~Ansart, and
  M.~Hemati.
\newblock A new heat transfer fluid for concentrating solar systems: {P}article
  flow in tubes.
\newblock {\em Energy Procedia}, 49:617--626, 2014.
\newblock Proceedings of the SolarPACES 2013 International Conference.

\bibitem{FLAMANT2013567}
Gilles Flamant, Daniel Gauthier, Hadrien Benoit, Jean-Louis Sans, Roger Garcia,
  Benjamin Boissière, Renaud Ansart, and Mehrdji Hemati.
\newblock Dense suspension of solid particles as a new heat transfer fluid for
  concentrated solar thermal plants: On-sun proof of concept.
\newblock {\em Chemical Engineering Science}, 102:567--576, 2013.

\bibitem{glowinski1}
R.~Glowinski, T.-W. Pan, T.I. Hesla, and D.D. Joseph.
\newblock A distributed {L}agrange multiplier/fictitious domain method for
  particulate flows.
\newblock {\em International Journal of Multiphase Flow}, 25(5):755–794, Aug
  1999.

\bibitem{glowinski2}
R.~Glowinski, T.W. Pan, T.I. Hesla, D.D. Joseph, and J.~Périaux.
\newblock A fictitious domain approach to the direct numerical simulation of
  incompressible viscous flow past moving rigid bodies: Application to
  particulate flow.
\newblock {\em Journal of Computational Physics}, 169(2):363–426, May 2001.

\bibitem{hwangbo_lee_hutter_2018}
Jemin Hwangbo, Joonho Lee, and Marco Hutter.
\newblock Per-contact iteration method for solving contact dynamics.
\newblock {\em IEEE Robotics and Automation Letters}, 3(2):895--902, 2018.

\bibitem{Kroupa_Kosek_Soos_Vonka_2016}
Martin Kroupa, Juraj Kosek, Miroslav Soos, and Michal Vonka.
\newblock Utilizing the discrete element method for the modeling of viscosity
  in concentrated suspensions.
\newblock {\em Langmuir}, 32(33):8451–8460, 2016.

\bibitem{lefebvre2005apparent}
Aline Lefebvre and Bertrand Maury.
\newblock Apparent viscosity of a mixture of a {N}ewtonian fluid and
  interacting particles.
\newblock {\em Comptes Rendus M{\'e}canique}, 333(12):923--933, 2005.

\bibitem{patankar2000}
N.A. Patankar, P.~Singh, D.D. Joseph, R.~Glowinski, and T.-W. Pan.
\newblock A new formulation of the distributed {L}agrange
  multiplier/{F}ictitious domain method for particulate flows.
\newblock {\em International Journal of Multiphase Flow}, 26(9):1509–1524,
  Sep 2000.

\bibitem{faucris.109633524}
Rodolphe Prignitz and Eberhard Bänsch.
\newblock {Numerical} simulation of suspension induced rheology.
\newblock {\em Kybernetika}, 46:281--293, 2010.

\bibitem{stewart_trinkle_1996}
D.~E. Stewart and J.~C. Trinkle.
\newblock An implicit time-stepping scheme for rigid body dynamics with
  inelastic collisions and {C}oulomb friction.
\newblock {\em International Journal for Numerical Methods in Engineering},
  39(15):2673--2691, 1996.

\bibitem{PhysRevLett.129.078001}
Franco Tapia, Mie Ichihara, Olivier Pouliquen, and \'Elisabeth Guazzelli.
\newblock Viscous to inertial transition in dense granular suspension.
\newblock {\em Phys. Rev. Lett.}, 129:078001, Aug 2022.

\bibitem{PhysRevLett.109.118305}
Martin Trulsson, Bruno Andreotti, and Philippe Claudin.
\newblock Transition from the viscous to inertial regime in dense suspensions.
\newblock {\em Phys. Rev. Lett.}, 109:118305, Sep 2012.

\bibitem{Turek1999}
Stefan Turek.
\newblock {\em Efficient solvers for incompressible flow problems: An
  algorithmic and computational approach}.
\newblock Springer, 1999.

\bibitem{afc3}
Stefan Turek and Dmitri Kuzmin.
\newblock Algebraic flux correction {III: I}ncompressible flow problems.
\newblock In Dmitri Kuzmin, Stefan Turek, and Rainald Löhner, editors, {\em
  Flux-Corrected Transport}, pages 239--297. Springer, 2 edition, 2012.

\bibitem{Vinogradova_Yakubov_2003}
Olga~I. Vinogradova and Gleb~E. Yakubov.
\newblock Dynamic effects on force measurements. 2. {L}ubrication and the
  atomic force microscope.
\newblock {\em Langmuir}, 19(4):1227–1234, Jan 2003.

\bibitem{WanTurek2006a}
Decheng Wan and Stefan Turek.
\newblock Direct numerical simulation of particulate flow via multigrid {FEM}
  techniques and the fictitious boundary method.
\newblock {\em International Journal for Numerical Methods in Fluids},
  51(5):531–566, Dec 2005.

\bibitem{WanTurek2006b}
Decheng Wan and Stefan Turek.
\newblock Modeling of liquid-solid flows with large number of moving particles
  by multigrid fictitious boundary method.
\newblock {\em Journal of Hydrodynamics}, 18(S1):93–100, Feb 2006.

\bibitem{WanTurek2007a}
Decheng Wan and Stefan Turek.
\newblock An efficient multigrid-{FEM} method for the simulation of
  solid–liquid two phase flows.
\newblock {\em Journal of Computational and Applied Mathematics},
  203(2):561–580, Jun 2007.

\bibitem{wang2015warmstartingprojectedgaussseidel}
Da~Wang, Martin Servin, and Tomas Berglund.
\newblock Warm starting the projected {G}auss-{S}eidel algorithm for granular
  matter simulation.
\newblock {\em Computational Particle Mechanics}, 3:43--52, 2016.

\bibitem{YEO20102401}
Kyongmin Yeo and Martin~R. Maxey.
\newblock Simulation of concentrated suspensions using the force-coupling
  method.
\newblock {\em Journal of Computational Physics}, 229(6):2401--2421, 2010.

\end{thebibliography}
\end{document}